\documentclass[useAMS,usenatbib,usegraphicx,printer]{mn2e}
\usepackage{footmisc}
\title{A long-lived horseshoe companion to the Earth}
\author[Christou \& Asher]{A.~A.~Christou$^{1}$\thanks{E-mail: aac@arm.ac.uk (AAC)}, D.~J.~Asher$^1$ \\
$^{1}$Armagh Observatory, 
College Hill, Armagh BT61 9DG, 
Northern Ireland, UK}
\begin{document}

\date{Accepted 2011 February 23; Received 2011 February 9; in original form 2010 December 18}

\pagerange{\pageref{firstpage}--\pageref{lastpage}} \pubyear{2011}

\maketitle

\label{firstpage}


\begin{abstract}
We present a dynamical investigation of a newly found asteroid, 2010 SO16, and the discovery that it is a horseshoe companion of the Earth.
The object's absolute magnitude ($H=20.7$) makes this the largest object of its type known to-date.
By carrying out numerical integrations of dynamical clones, we find that (a) its status as a horseshoe is secure given the current
accuracy of its ephemeris, and (b) the time spent in horseshoe libration with the Earth is several times $10^{5}$ yr, 
two orders of magnitude longer than determined for other horseshoe asteroids of the Earth. Further, using a model based on Hill's 
approximation to the three-body problem, we show that, apart from the low eccentricity which prevents close encounters with other planets 
or the Earth itself, its stability can be attributed to the value of its Jacobi constant  far from the regime that allows transitions 
into other coorbital modes or escape from the resonance altogether. We provide evidence that the eventual escape of the asteroid from horseshoe 
libration is caused by the action of planetary secular perturbations and the stochastic evolution of the eccentricity. 
The questions of its origin and the existence of as-yet-undiscovered co-orbital companions of the Earth are discussed.
\end{abstract}

\begin{keywords}
celestial mechanics -- minor planets, asteroids: general -- minor planets, asteroids: individual: 2010 SO16
\end{keywords}


\section{Introduction}
\label{intro}
Horseshoes and tadpoles comprise the two classical solutions to the special case of coorbital motion 
in the context of the restricted three-body problem. Both types of motion have been studied extensively, 
both analytically and numerically, partly due to their wide applicability to quasi-stable configurations 
in the real solar system.
Horseshoe-type orbits appear to be less prevalent than tadpoles, owing perhaps to their different
stability properties. An often-cited example of objects horseshoeing with each other are the 
Saturnian satellites Janus and Epimetheus. In addition, numerical integrations of 
the motion of Near-Earth Asteroids (NEAs) 54509 YORP and 2002 AA29 show that they are currently horseshoeing with
the Earth \citep{con02,bra04}. Confirmation of the status of NEA 2001 GO2 as the third Earth 
horseshoe will have to await further refinement of its orbit \citep{bra04}. The 
libration periods and lifetimes of these Earth horseshoes are typically a few hundred 
and several thousand years respectively. 

Here we demonstrate the horseshoe dynamics of a newly discovered NEA, 2010 SO16. We find that, in contrast
to the previous cases, the lifetime of horseshoe libration between 2010 SO16 and the Earth is 
substantially higher, of order $10^{5}$ years. We attribute this to two factors: its
low eccentricity, moderate inclination orbit and its energy state deep into
the horseshoe regime. We discuss the issue of its origin and briefly comment on the possible
existence of additional objects of this type.  
\begin{table}
  \caption{Orbital elements of 2010 SO16 at JD2455400.5.}
  \label{elements}  
  \begin{center}
  \begin{tabular}{lcr}
\hline 
Element  &   Value & 1-$\sigma$ uncert. \\ \hline  
a  (au)  &   1.00039  & 9.961e-6 \\  
e        &   0.075188 & 3.284e-6 \\
$i$  (deg)      &   14.536  & 0.0008793 \\
$\omega$ (deg)  &   108.283 & 0.003205 \\
$\Omega$  (deg) &  40.523 & 0.001306 \\
M  (deg)      &  137.831 & 0.004636 \\ \hline
  \end{tabular}
  \end{center} 
\end{table}

\section{The Asteroid}
\label{asteroid}
2010 SO16, hereafter referred to as ``SO16'', was discovered on 2010 September 17 by the WISE 
Earth-orbiting observatory (Obs.~Code C51)
and subsequently followed up by ground-based telescopes. As of 2010/12/03, 
its orbit has been determined by 40 observations spanning a data arc of 75 days (MPEC 2010-X02). The
orbital elements and their 1-$\sigma$ statistical uncertainties as given by the NEODYS orbital 
information service at that date (http://newton.dm.unipi.it/neodys/) are shown in Table~\ref{elements}.  
Its Minimum Orbit Intersection Distance (MOID) with respect to the Earth is 0.027 au and its absolute 
magnitude $H$ is $20.7$ implying a diameter of 200-400 m depending on the object's unknown albedo.
\begin{figure}
\includegraphics[width=84mm]{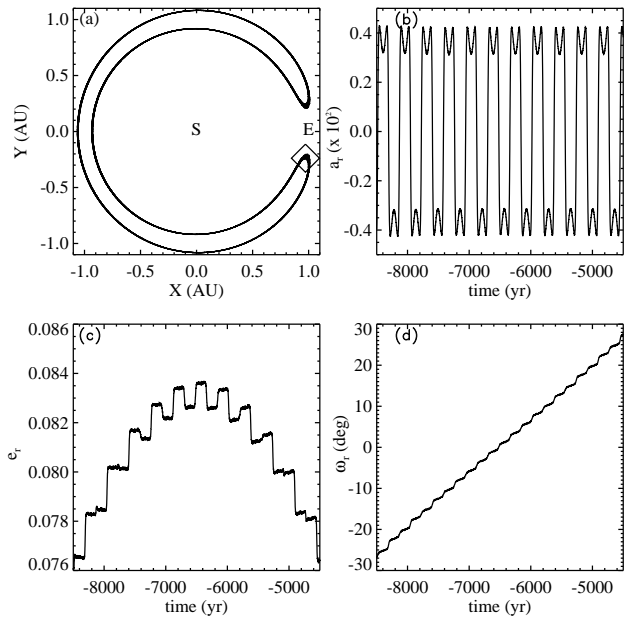}
\caption{Panel (a): The guiding centre trajectory of NEO 2010 SO16 in a cartesian ecliptic heliocentric frame 
co-rotating with the Earth from an integration of the asteroid's nominal orbit solution 
(Table \ref{elements}) and for an interval of $2\times 10^{5}$ yr centred on 
2000 Jan 1.5 UT. The radial extent of the horseshoe has been exaggerated by a factor of 20 for clarity. 
The positions of the Sun and the Earth are denoted by the letters `S' and `E' 
respectively. The diamond indicates the position of the asteroid at 2010 Jan 1.\mbox{.5} UT.
Panel (b): Relative semimajor axis showing the alternation between two values, one interior and the other exterior to 
the Earth's, typical of horseshoe libration.
Panels (c) and (d): Relative eccentricity and argument of pericentre respectively. Note the stepwise behaviour owing to the 
object's encounter with the Earth every $\sim 175$ yr.}
\label{hshort}
\end{figure}

\section{Verifying the object's dynamical state}
\label{clones}
Following astrometry of SO16 by the Spacewatch II telescope (Obs.~Code 291) on 18 Nov, and while the obs.~arc length was 62 days,
the nominal orbit solution as given by the NEODYS ephemeris information system had a semimajor axis ($a$) uncertainty of 
$\sigma_{\rm a}=1.6 \times 10^{-5}$ au while the difference between the value of this orbital element and the corresponding value
for the Earth was $\Delta a \sim 4 \times 10^{-4}$ au. Both are small compared to the range of possible values for $\Delta a$ 
that allow co-orbital motion (${\Delta a}_{\rm max} \sim 10^{-2}$ au) suggesting that SO16 is a 1:1 resonant librator.

We numerically integrated this nominal orbit, as well as the one subsequently published following additional observations obtained on the 1st December
(Table~\ref{elements}), for $\pm10^{5}$ yr using the RADAU algorithm implemented within the MERCURY integration package 
\citep{cha99}. The model of the solar system used in the integrations included all eight major planets 
\cite[ initial state vectors for which were taken from the HORIZONS
ephemeris service;][]{gio96} with the gravity of the Moon treated by adding its mass to that of the Earth.
In both integrations, the object persisted as a horseshoe librator of the Earth. Figure~\ref{hshort} (a) shows 
the cartesian motion of the asteroid for a period of $2 \times 10^{5}$ yr, averaged over one orbital period, on the ecliptic plane. 
The opening angle of the horseshoe is $\sim 25\degr$ and the half-amplitude of variation in $a$ is $\sim 4 \times 10^{-3}$ au.
The period of the horseshoe libration is $\sim 350$ yr. SO16 is currently at the turning point of the horseshoe that 
is lagging behind the Earth in its orbital motion, approaching it at 0.13 au in mid-May every year and at $<0.2$ au until 2016. 
It will remain as an evening object in the sky for several decades to come.
 
To test this result further, and explore the likely lifetime of the object in the horseshoe state, 
we integrated dynamical clones of SO16. Starting conditions for these were generated by varying 
the asteroid's semimajor axis $a$, mean anomaly $M$, eccentricity $e$ and argument of pericentre $\omega$ 
by increments equal to their 1-$\sigma$ ($a$) and 1.5-$\sigma$ ($M$, $e$, $\omega$) ephemeris uncertainties respectively. 
Nine different values were used for the semimajor axis and three for the other elements (inclusive of the nominal values), 
resulting in 243 clones which were integrated with MERCURY for $\pm 10^{5}$ yr with respect to the present. 
All clones remained in horseshoe libration with respect to the Earth.  
\section{Analysis of the Dynamics}      
\label{dynamics}   
\subsection{Theoretical Framework}
\label{theory}
The existence of horseshoe trajectories within the circular restricted three body problem was
first demonstrated by \citet{bro11}. The realisation that two recently discovered satellites of Saturn, Janus and Epimetheus, 
were engaged in mutual horseshoe libration \citep{dm81b}, provided new impetus for theoretical modelling of the dynamics.

\citet{dm81a} investigated quasi-circular symmetric horseshoe orbits. In the symmetric regime, the initial separation in semimajor axis 
between the orbits (the impact parameter) and the eccentricity are adiabatic invariants of the motion, recovering their initial values following 
two consecutive encounters. In that work, it was also demonstrated numerically that the assumption of symmetry is valid up to 
$|a_{\rm 0}|\simeq 0.74 \epsilon $. Here, $\epsilon$ is the mass parameter, a fundamental scaling constant in Hill's 
approximation -- where the indirect perturbation is ignored and only the interaction potential is considered during 
the encounter -- equal to ${\left(\mu /3 \right)}^{1/3}$  where $\mu$ is the mass ratio between the secondary (or secondaries) and the primary. 
The quantity $a_{0}$ is the relative semimajor axis far from encounter, equivalent to the Jacobi constant for planar, circular orbits. 
As the symmetry is not exactly preserved, the horseshoe configuration has a finite lifetime of $T/\mu^{5/3}$ where $T$ is the orbital period 
of the secondary (eg for the Earth, this expression evaluates to $\sim 1.6 \times 10^{9}$ yr). 
\citet{hp86} formally demonstrated that the eccentricity receives an impulse of order $\exp{\left(- 8 \pi \mu/3 a^{3}_{0}\right)}$ 
during an encounter and hence is an adiabatic invariant of the motion \cite[see also][]{nam99}.  For small to moderate 
eccentricities ($\la \epsilon$) the assumption of symmetry is valid if $\phi_{\rm min} \gg a_{\rm 0}$ where   
$\phi_{\rm min}$, related to the impact parameter by
\begin{equation}
{\phi}_{\rm min} = 8 \mu / 3 a^{2}_{0}\mbox{,}
\label{encounter}
\end{equation}
is the angular separation between the two bodies at encounter.

In the context of Hill's three-body problem \citep{hil78}, the evolution of the asteroid's orbit can be studied by averaging over 
the fast epicyclic motion \citep{dm81a,hp86,nam99}. \citet{nam99} derived an expression that describes the slow 
variation of the guiding centre:
\begin{eqnarray}
a^{2}_{\rm r}&=&a^{2}_{0} - \frac{4 \mu}{3 \pi}\int_{-\pi}^{\pi}\left[{\left(e_{\rm r}\cos u + a_{\rm r}\right)}^{2} + 4 {\left(l_{\rm r} - e_{\rm r} \sin u \right)}^{2} \right. \nonumber\\
  &  & \left. + I^{2}_{\rm r} \sin {\left( u + \omega_{\rm r} \right)}^{2} \right]^{-1/2} du \label{gc1}
\end{eqnarray} 
where the subscript ``r'' denotes the relative elements as defined in that work  and $l_{\rm r} = \phi/2$.  The constant $a_{\rm 0}$ is related to 
the Jacobi constant $H_{\rm r}$ and the conserved Poincar\'{e} action $K_{r} = 1 - \sqrt{ 1 - e^{2}_{\rm r} \cos I_{\rm r}}$: 
\begin{equation}
a^{2}_{\rm 0} = \frac{8}{3} \left( K_{\rm r} - H_{\rm r}\right)
\label{a0def}
\end{equation}
assuming a planetary mean motion of unity.

In our case, $a_{\rm r}  \ll e_{\rm r}$ 
and we can write: 
\begin{eqnarray}
a^{2}_{\rm r}&=&a^{2}_{0} - \frac{4 \mu}{3 \pi e_{\rm r}}\int_{-\pi}^{\pi}\left[\cos^{2}u + 4 {\left(\bar{l} - \sin u \right)}^{2} \right. \nonumber\\
  &  & \left. + \left( \bar{K} - 1 \right) \sin {\left( u + \omega_{\rm r} \right)}^{2} \right]^{-1/2} du \label{gc2}
\end{eqnarray}
where  $\bar{K} = 1 + I^{2}_{\rm r}/e^{2}_{\rm r}$ and $\bar{l}=l_{\rm r}/e_{\rm r}$.
The second term on the right-hand-side of this equation is the ponderomotive potential $S$ scaled by $8 \mu / 3  e_{\rm r}$ \cite[Eq.~29 of][]{nam99}. 
It is $\pi$-periodic in $\omega_r$ and symmetric with respect to $\bar{l}=0$.
 It was also shown in that paper that averaging the co-orbital potential over the phase $\bar{l}$ (or $l_{\rm r}$) 
allows one to obtain analytically the secular evolution of $e_{\rm r}$, $I_{\rm r}$, $\omega_{\rm r}$ and 
$\Omega_{\rm r}$ if the orbits are non-collisional. In fact, the regularity of the secular dynamics persists at large 
eccentricity and inclination if Hill's equations are replaced with those of the circular restricted three body problem.   

\subsection{Short Term Evolution}
\label{short}
In the case of SO16, it is expected that the orbital elements will evolve in the short term due to the coorbital resonance and in the long term 
due to secular perturbations. To understand the dynamical evolution of SO16 we must first test the assumptions under which the results of the 
previous section are valid. 
The condition that $\phi_{\rm min} \gg a_{\rm 0}$ is satisfied as $|a_{\rm 0}| \leq \epsilon$, however $e_{\rm r}$ 
and $I_{\rm r}$ are large relative to the mass parameter.   

In Fig.~\ref{hshort} we show the variation of $a_{\rm r}$, $e_{\rm r}$ and $\omega_{\rm r}$ for several coorbital cycles. 
We observe impulsive stepwise changes in $e_{\rm r}$ at the same time that $a_{\rm r}$ changes sign upon encounter with the Earth. 
These step changes cancel each other out over a full libration cycle (two consecutive encounters) so that the original eccentricity 
is recovered. This behaviour of impulsive changes, also seen in the plot for $\omega_{\rm r}$ as well as the evolution of 
$I_{\rm r}$ and $\Omega_{\rm r}$ (not shown here), is similar to that observed in the integrations of \citet{dm81a} for 
symmetric orbits. It is superposed on a slower, gradual evolution of these orbital elements. These observations indicate that SO16 currently 
occupies the symmetric regime of coorbital motion as discussed in Section~\ref{theory} and that the secular motion is regular at these timescales.    
\begin{figure}
\includegraphics[width=84mm]{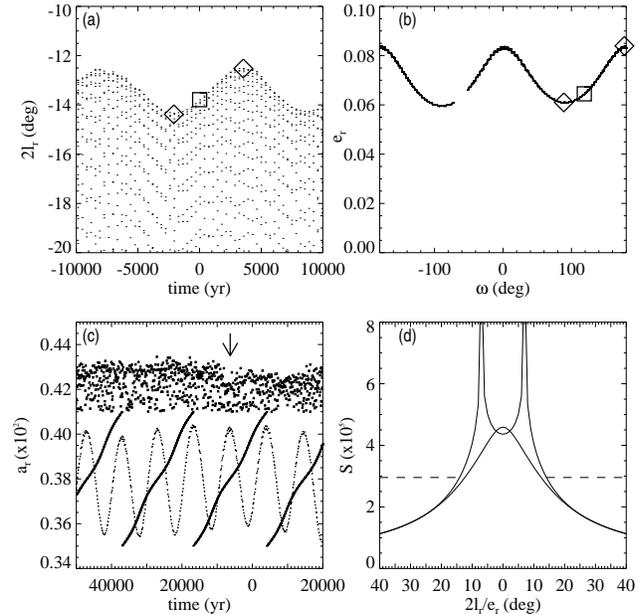}
\caption{Evolution of the nominal orbit of 2010 SO16 (Table~\ref{elements}) for timescales of order $10^{4}$ yr 
showing key features of the dynamics.
Panel (a): The Earth-trailing branch of the relative mean longitude $2 l_{\rm r}$ between SO16 and the planet. 
Note the $\sim 10^{4}$ yr periodicity. The rectangle indicates the value of $2 l_{\rm r}$ at JD 2455197.5.
Panel (b): Evolution of the relative eccentricity ($e_{\rm r}$)  as a function of the relative argument of pericentre ($\omega_{\rm r}$).
The diamonds in this and the previous panel indicate the extreme states referred to in Section~\ref{short}. 
Panel (c): The evolution of the libration width of  the relative semimajor axis $a_{\rm r}$ 
(for clarity, only values $\geq +0.41 \epsilon$ are shown) through a jump, indicated by the arrow, at $t = -6.3 \times 10^{3}$ yr. 
The time evolution of $e_{\rm r}$ and $\omega_{\rm r}$ is superimposed on the plot as the oscillating and slanted curves 
respectively. 
Panel (d): The form of the scaled ponderomotive potential $\left(8 \mu / 3  e_{\rm r}\right) S$ corresponding to the two snapshots 
in $e_{\rm r}$-$\omega_{\rm r}$ evolution highlighted in panel (b) and Section~\ref{short}. 
The horizontal dashed line indicates the current state of SO16's horseshoe trajectory as quantified by the constant $a^{2}_{0}$.
This is to be compared with the two values of the extremum at $\bar{l}=0$ referred to in the text.}
\label{hlong}
\end{figure}

Therefore, the formulation presented in Section~\ref{theory} may be used as a starting point in our effort to explain the 
pertinent characteristics of 2010 SO16's long-term orbital evolution.
Note that, as the Earth's orbital eccentricity becomes significant within the timescales considered in our integrations
\citep{qtd91}, its contribution to $e_{\rm r}$ cannot be neglected.   

By evaluating Eq.~\ref{gc2} at JD 2455197.5  (2010 Jan 1.0 UT) when $l_{\rm r}=-13\degr.8/2$, 
$e_{\rm r}=0.0645$, $I_{\rm r}=14\degr.5$, $\omega_{\rm r}=118\degr.8$ and setting $a_{\rm r}=0$, 
we find $a_{0}=0.544\epsilon$. The corresponding angle from Eq.~\ref{encounter} is $15\degr.5$, in good agreement given 
that the true orbit has a non-negligible eccentricity and inclination.

In panel (a) of Fig.~\ref{hlong} we observe an oscillation of ${\phi}_{\rm min}$ with a period of $\sim 10^{4}$ yr and between $14\degr.3$ 
(at $t=-2 \times 10^{3}$ yr) and $12\degr.5$ (at $t=+3 \times 10^{3}$ yr). This is due to the contribution of $e_{\rm r}$ in Eq.~\ref{gc2} 
 which oscillates with the same period. This element, together with $\omega_r$, form a conjugate pair of variables which, as 
our integrations show, oscillate between two states; one where $e_{\rm r}=e_{\rm r,max}=0.083$ and $\omega_{\rm r} = k \pi$ and the 
other where $e_{\rm r}=e_{\rm r, min}=0.061$ and $\omega_{\rm r}= k \pi + \pi/2$ (panel (b) of Fig.~\ref{hlong}). 
This cycle,  hereafter referred to as the eccentricity cycle, by itself defines the object's closest possible distance to 
the Earth as $\simeq$ 0.12 au.  Two such cycles are completed for every full circulation of $\omega_{\rm r}$. 
This behaviour is characteristic of the horseshoe state \cite[cf Fig.~12 of][]{nam99} and demonstrates the regularity of the secular evolution 
of this asteroid's orbit over at least $2 \times 10^{4}$ yr or $\simeq 10^{2}$ horseshoe encounters with the Earth.
 Over the same period, the inclination $I_{\rm r}$ and action $K_{r}$ remain essentially constant and equal to 
$\sim 14\degr$ and $\sim 0.0340$. If we substitute these values in Eq.~\ref{gc2} we find ${\phi}_{\rm min}=14\degr.1$ 
and ${\phi}_{\rm min}=12\degr.4$ for the low-e and high-e states respectively, in excellent agreement with the simulations.  

The horseshoe trajectory of the guiding centre appears to widen at $t \simeq -6.3 \times 10^{3}$ yr (panel (c) of Fig.~\ref{hlong}). 
This coincides with $e_{\rm r}$ reaching its maximum. However, similar behaviour is not observed at the next $e_{\rm r}$ maximum 
($t=+3 \times 10^{3}$ yr) even though the same value is attained.
Inspection of our simulations shows that this increase takes place for 10-15\% of our clones, indicating that it is not deterministic.
Its size is ${\Delta a}_{\rm r} \simeq 0.01 \epsilon$. Nothing similar is observed in the time history of the other elements.
Its impulsive nature indicates that this asteroid's orbit evolves towards the boundary of the symmetric horseshoe regime over time. 

The ultimate fate of the horseshoe depends on the relative evolution of $a_{\rm 0}$ and the ponderomotive potential $S$. 
Since $e_{\rm r} < I_{\rm r}$ the extremum of the scaled ponderomotive potential at $\bar{l}=0$ changes 
from a maximum to a minimum when $\omega_{\rm r}> \omega_{0}(\bar{K})$ \citep{nam99} (see panel (d) of Fig.~\ref{hlong}). 
This threshold evaluates to $30\degr$ for the average $e_{\rm r}$ and $I_{\rm r}$ of this asteroid's orbit. 
In any case, the horseshoe will transition to, or form a compound orbit with, a quasi-satellite orbit or a passing orbit when 
$a^{2}_{0} \simeq (8 \mu / 3 e_{\rm r}) S $. For example, and for the $e_{\rm r}$-$\omega_{\rm r}$ behaviour observed in the integrations, 
the value of the extremum varies from ${(0.661 \epsilon)}^{2}$ to ${(0.678 \epsilon)}^{2}$ (see panel (d) of Fig.~\ref{hlong}). 
Hence, an increase in $a_{0}$ of $ \ga 0.12 \epsilon$ is required in order for the horseshoe to be disrupted. 
%
%

\subsection{Long Term Evolution}
\label{long}
\begin{figure}
\includegraphics[width=82mm,height=95mm]{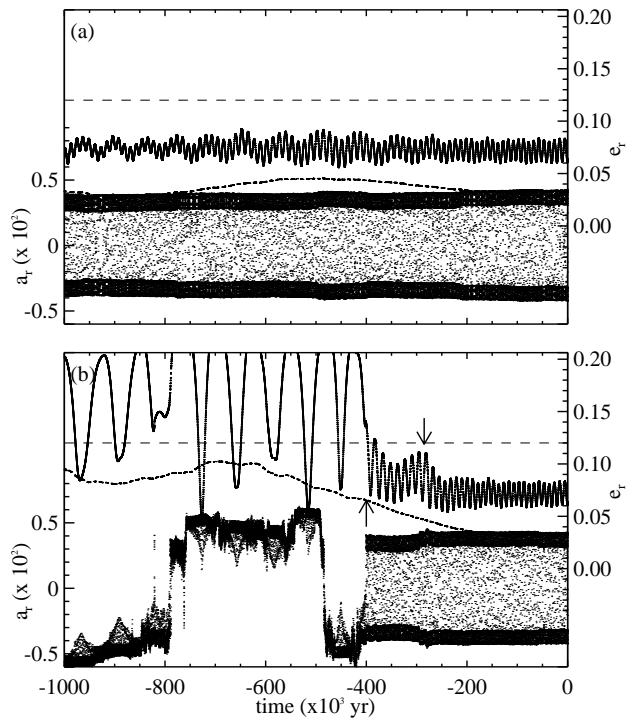}
\caption{Evolution of two dynamical clones of 2010 SO16 for ${10}^{6}$ yr in the past. Panel (a) shows a clone that persists as 
a horseshoe of the Earth for this period. Panel (b) shows a horseshoe that is disrupted at $t = -4 \times 10^{5}$yr following 
two impulsive changes in $e_{\rm r}$ (indicated by the arrows). The dashed horizontal line indicates the critical value of the 
eccentricity ($\sim 0.12$) as discussed in the text. The evolution of the Poincar\'{e} action $K_{\rm r}$, in the same scale as 
$e_{\rm r}$, is denoted by the dashed curve.}
\label{hlonger}
\end{figure}

To explore further the question of the horseshoe's likely survival lifetime and the mechanisms which might limit it, we generated additional 
clones by sampling the semimajor axis with a step of 0.1-$\sigma$ out to $\pm 1.4\sigma$ and then at $\pm 2$, $\pm 3$ and $\pm 4 \sigma$. 
This new set of 35 clones was integrated for $\pm 2 \times 10^{6}$ yr. We regard the integrations in the past and in the future as separate trials. 
Out of this set of 70 trials, 50 persisted as horseshoes for $2-5 \times 10^{5}$ yr, 4 for $< 2 \times 10^{5}$ yr and the remaining 16 
for $> 5 \times 10^{5}$. Eight clones persisted for $> 10^{6}$ yr while two remained in horseshoe libration until the end of the integration. 
The shortest lifetime observed was $1.2 \times 10^{5}$ yr. 
In Fig.~\ref{hlonger} we show two typical cases of a clone that is stable and one that escapes 
within $10^{6}$ yr. In panel (a), the width of the libration in $a_{\rm r}$ remains essentially constant over this period. 
At the same time, the behaviour of the eccentricity remains regular, staying well below the collision threshold (dashed horizontal line; 
see below for explanation) despite subtle changes (eg the $\sim 5 \times 10^{4}$ yr modulation that sets in towards the middle 
of the integration). Interestingly, $K_{\rm r}$ also evolves in a regular fashion over this period, reaching a maximum value of 
$\sim 0.05$ near the middle of the time interval. 
The timescale of variation implies that this evolution is due to secular forcing by the planets, acting to change the asteroid's 
(and the Earth's) orbital elements over $> 5 \times 10^{4}$ yr timescales.
  
On the other hand, the behaviour of $a_{\rm r}$ and $e_{\rm r}$ of the clone shown in panel (b) changes markedly at 
$t= -2.8 \times 10^{5}$ yr and $t= -4.0 \times 10^{5}$ yr. In the first instance, we observe an increase of both 
the average eccentricity and the amplitude of the secular cycle while the width of the horseshoe decreases slightly by 
$ 0.04 \epsilon$. In the second instance, a further increase of the amplitude of the eccentricity cycle leads to 
a transition into a passing orbit.
In both of these instances, the change occurs over 1-2 eccentricity cycles so no clear correlation with particular phases of the 
cycle - of the sort evident in Fig.~\ref{hlong}(c) - can be made. Establishing such a correlation would require a thorough statistical analysis 
that is outside the scope of the paper.
However, $K_{r}$ evolves smoothly through the change (ie no jumps are observed), suggesting that they are dynamical features of the 
restricted Sun-Earth-Asteroid problem. On the other hand, the timescale over which these events occur, similar to that governing the evolution 
of $K_{r}$, implies that planetary secular perturbations also play a role.
The transition itself occurs because the maximum of the scaled potential $S$ is reduced to the point of reaching
parity with $a^{2}_{\rm 0}$. In physical terms, the eccentricity becomes large enough 
($\sim 0.12$; indicated by the dashed horizontal line) to allow close encounters ($<0.01$ au) with Earth while the inclination 
acts to stabilise the transition. Afterwards, the eccentricity becomes high enough to allow close encounters with Venus and the evolution
is similar to that of other high-$e$, high-$I$ coorbitals of the Earth \citep{chr00}.   
     
%
\section{Comparison with other Earth horseshoe asteroids}  
\begin{table*}
  \caption{Known Earth horseshoe asteroids. Data for 2002 AA29 and 2001 GO2 were taken from \citet{con02} and \citet{bra04}. 
Information for 3753 Cruithne \citep{wim98} is also listed as this object is the archetype of the class of transient coorbitals of the Earth. 
Parameters of 54509's horseshoe libration were determined by numerical integration of 9 dynamical clones of that asteroid. Additional information was taken from \citet{mn03}.}
  \label{other}  
  \begin{center}
  \begin{tabular}{lrrrrrrrr}
\hline
                   &     $a_{\rm r,max}$     &    &  $I$     &  Obs.~Data Arc${}^{\rm a}$ & Lifetime     &  $ 2 \phi_{\rm min}$ & Libration Period &  \\ 
Designation   &    ($\epsilon$)  &  $e$   & ($\degr$)  &  (days)   & (yr)      &   ($\degr$)     &    (yr)          & $H$\\ \hline
2010 SO16  &   0.42   &   0.07 &  14.5  &  75   &     $>$120,000      &      25      &    350      &     20.7\\
2002 AA29  &    0.70   &    0.01 &  10.7 &   735  &     2000    &        $\sim$8   &       190     &      24.1\\
2001 GO2${}^{b}$ &    0.68   &    0.17 &  4.6  &    5    &      $\leq$ 2500    &         36    &      200      &     23.5\\
54509 YORP &    0.59   &    0.23 &  1.6 &   1100    &   $\leq$ 2200    &     54      &   200     &      22.6   \\  
3753 Cruithne${}^{c}$   &       0.23   &     0.51 &  19.8 &   4200   &     5500     &       N/A    &     770    &       15.5   \\ \hline
${}^{a}$ At time of demonstration of horseshoe state\\
${}^{b}$ Obs. arc length 5 days \\
${}^{c}$ Compound orbit
\end{tabular}
\end{center} 
\end{table*}
There are currently 3 near-Earth asteroids known to follow horseshoe trajectories with respect to the Earth \cite[Table~\ref{other}; also][]{bra04}.
Of those, 2002 AA29 is the most similar to 2010 SO16, mainly due to its low eccentricity and moderate inclination. 
Its value of $H$ implies a likely size of a few tens of metres. Hence, it is an order of magnitude smaller than SO16.

The orbital eccentricities of 54509 YORP and 2001 GO2 allow close encounters with the Earth. These do not necessarily eject the 
asteroids from the co-orbital resonance but instead they effect their transition into another mode of libration or circulation.
This state is shared by other objects in so-called compound and/or transition orbits such as 3753 Cruithne, 
the first of its class to be recognised as such \citep{wim97,wim98,nam99,ncm99,chr00}. The opening angles 
$\mbox{\boldmath$2 \phi_{\rm min}$}$ of GO2 and 54509 are wider than those of SO16 implying an energy state deeper into the horseshoe domain. 
However, the contribution of their higher eccentricities in Eq.~\ref{gc2} places them closer to the transition regime
(see panel (d) of Fig~\ref{hlong}) than SO16 and accounts, at least in part, for their shorter lifetimes - a few times $10^{3}$ yr - as horseshoe companions of the Earth. 

\section{Discussion}
\label{discussion}
The existence of this long-lived horseshoe raises the twin questions of its origin and whether objects in similar orbits are yet to be found.
The object's Earth-like orbit makes a direct or indirect origin in the main belt an unlikely, although not impossible, proposition \citep{bot00,bw08}.
Another plausible source is the Earth-Moon system. \citet{mn03} have suggested that 54509 may have originated within the Earth-Moon system. 
SO16's current orbit does not provide such a direct dynamical pathway to the Earth-Moon system as in that case, although the situation is 
likely to change within a timescale of several times $10^{5}$ yr. A third possibility is that the object originated
near the Asteroid-Earth-Sun L4 or L5 equilibrium point as a tadpole librator. \citet{te00} showed that Earth tadpoles can persist for 
$5 \times 10^{7}$ yr in both high ($24\degr < I < 34\degr$) and low ($I < 16 \degr$) inclination orbits and at typical eccentricities of $\sim0.06$. 
Extrapolating their results to $5 \times 10^{9}$ yr timescales, they found it conceivable that a small fraction of a postulated initial 
population of Earth tadpoles can survive to the present. In order for this scenario to apply to SO16, it must have been a tadpole 
until very recently. However, the work of \citeauthor{te00} does not take into account the action of the Yarkovsky effect, 
which mobilises the semimajor axes of NEAs at typical rates of $10^{-9}$ au $\mbox{yr}^{-1}$ (Chesley et al, Ast.~Comets Met.~Conf., 2008). 
As the radial half-width of Earth's tadpole region is $\simeq \sqrt{8 \mu / 3} \sim 2.8 \times 10^{-3}$ au \citep{md99}, it may be difficult 
for SO16 to survive as a tadpole companion of the Earth for more than a few times $10^{6}$ yr. On the other hand, a study of the stability of 
Trojans of Mars has shown that the Yarkovsky acceleration does not necessarily destabilise large Trojan companions of the planets \citep{sch05} 
and this result may also apply here. In any case, observational determination of the asteroid's basic properties - size, spectral class 
and spin state -  will be extremely useful in clarifying the situation. Subject to the successful determination of said parameters, SO16 
may be a suitable test target for the direct detection of the Yarkovsky acceleration as an object a few hundred metres across that makes 
frequent close encounters with the Earth during the next decade \citep{vok05}. 

Finally, regarding the possibility of existence of other bodies in similar orbits, the case for Earth-based reconnaissance 
of the Trojan regions of the Earth has already been made \citep{et00,wi00}. For relatively bright targets such as SO16, 
the main limiting factor is likely to be the minimum solar elongation that can be reached by an observational survey. As
typical horseshoe libration periods are of order hundreds of years, completing the census of large ($H<21$) terrestrial horseshoe or tadpole 
companions would likely require a space-based platform in a heliocentric orbit interior to the Earth's.    

\section*{Acknowledgments}
The authors would like to thank Ramon Brasser for his insightful comments which improved the manuscript.
Astronomical research at the Armagh Observatory is funded by the Northern 
Ireland Department of Culture, Arts and Leisure (DCAL).




\bsp

\label{lastpage}


\begin{thebibliography}{99}
\bibitem[\protect\citeauthoryear{Bottke et al}{2000}]{bot00}
Bottke W.~F., Jedicke R., Morbidelli A., Petit J.-M., Gladman B., 2000, Sci, 288, 2190.
\bibitem[\protect\citeauthoryear{Brasser \& Wiegert}{2008}]{bw08}
Brasser R., Wiegert P., 2008, MNRAS, 386, 2031.
\bibitem[\protect\citeauthoryear{Brasser et al}{2004}]{bra04}
Brasser R., Innanen K.~A., Connors M., Veillet C., Wiegert P., Mikkola S., Chodas P.~W., 2004,
Icarus, 171, 102.
\bibitem[\protect\citeauthoryear{Brown}{1911}]{bro11}
Brown E.~W., 1911, MNRAS, 71, 438.
\bibitem[\protect\citeauthoryear{Chambers}{1999}]{cha99}
Chambers J., 1999, MNRAS, 304, 793.
\bibitem[\protect\citeauthoryear{Christou}{2000}]{chr00}
Christou A.~A., 2000, Icarus, 144, 1.
\bibitem[\protect\citeauthoryear{Connors et al}{2002}]{con02}
Connors M., Chodas P., Mikkola S., Wiegert P., Veillet C., Innanen K., 2002,
Met.~Planet.~Sci.,~37, 1435.
\bibitem[\protect\citeauthoryear{Giorgini et al}{1996}]{gio96}
Giorgini J.~D., Yeomans D.~K., Chamberlin A.~B., Chodas P.~W., Jacobson R.~A., 
Keesey M.~S., Lieske J.~H., Ostro S.~J., Standish E.~M., Wimberly R.~N., 1996,
Bull.~Am.~Astron.~Soc., 28, 1158.
\bibitem[\protect\citeauthoryear{Dermott \& Murray}{1981a}]{dm81a}
Dermott S.~F., Murray C.~D., 1981a, Icarus, 48, 1.
\bibitem[\protect\citeauthoryear{Dermott \& Murray}{1981b}]{dm81b}
Dermott S.~F., Murray C.~D., 1981b, Icarus, 48, 12.
\bibitem[\protect\citeauthoryear{Evans \& Tabachnik}{2000}]{et00}
Evans, N.~W., Tabachnik S.~A., 2000, MNRAS, 319, 80.
\bibitem[\protect\citeauthoryear{H\'{e}non \& Petit}{1986}]{hp86}
H\'{e}non M., Petit J.-M., 1986, Cel.~Mech., 38, 67.
\bibitem[\protect\citeauthoryear{Hill}{1878}]{hil78}
Hill G.~W., 1878, Amer.~J.~Math., 1, 5.
\bibitem[\protect\citeauthoryear{Margot \& Nicholson}{2003}]{mn03}
Margot J.~L., Nicholson P.~D., 2003, BAAS, 35, 1039.
\bibitem[\protect\citeauthoryear{Murray \& Dermott}{1999}]{md99}
Murray C.~D., Dermott S.~F., 1999, Solar System Dynamics, Cambridge University Press, Cambridge.
\bibitem[\protect\citeauthoryear{Namouni}{1999}]{nam99}
Namouni F., 1999, Icarus, 137, 293.
\bibitem[\protect\citeauthoryear{Namouni et al}{1999}]{ncm99}
Namouni F., Christou A.~A., Murray C.~D., 1999, Phys.~Rev.~Lett., 83, 2506.
\bibitem[\protect\citeauthoryear{Quinn, Tremaine \& Duncan}{1991}]{qtd91}
Quinn T.~R., Tremaine S., Duncan M., 1991, AJ, 101, 2287.
\bibitem[\protect\citeauthoryear{Scholl et al}{2005}]{sch05}
Scholl H., Marzari F., Tricarico P., 2005, Icarus, 175, 379.
\bibitem[\protect\citeauthoryear{Tabachnik \& Evans}{2000}]{te00}
Tabachnik S.~A., Evans N.~W., 2000, MNRAS, 319, 63.
\bibitem[\protect\citeauthoryear{Vokrouhlick\'{y} et al.}{2005}]{vok05}
Vokrouhlick\'{y} D., Capek D., Chesley S.~R., Ostro S.~J., 2005, Icarus, 173, 166. 
\bibitem[\protect\citeauthoryear{Wiegert et al.}{1997}]{wim97}
Wiegert P.~A., Innanen K.~A., Mikkola S., 1997, Nat, 387, 685.
\bibitem[\protect\citeauthoryear{Wiegert et al.}{1998}]{wim98}
Wiegert P.~A., Innanen K.~A., Mikkola S., 1998, AJ, 115, 2604.
\bibitem[\protect\citeauthoryear{Wiegert et al.}{2000}]{wi00}
Wiegert P., Innanen K., Mikkola S., 2000, Icarus, 145, 33.
\end{thebibliography}
\end{document}